\documentclass[11pt]{article}

\usepackage{amsmath}
\usepackage{amssymb}
\usepackage{fullpage}
\usepackage{graphicx}
\usepackage{bm}
\usepackage{hyperref}
\usepackage{cite}
\usepackage{authblk}
\usepackage{color}

\PassOptionsToPackage{hyphens}{url}\usepackage{hyperref}

\title{\textbf{Evolution of cooperation in costly institutes}}
\date{}

\author{\textbf{Mohammad Salahshour}\thanks{\texttt{salahshour.mohammad@gmail.com}.}}
\affil{Max Planck Institute for Mathematics in the Sciences, Inselstrasse 22, D-04103, Leipzig, Germany}

\begin{document}
\maketitle

\begin{abstract}
We show that in a situation where individuals have a choice between a costly institute and a free institute to perform a collective action task, the existence of a participation cost promotes cooperation in the costly institute. Despite paying for a participation cost, costly cooperators, who join the costly institute and cooperate, can out-perform defectors, who predominantly join a free institute. This, not only promotes cooperation in the costly institute but also facilitates the evolution of cooperation in the free institute. A costly institute out-performs a free institute when the profitability of the collective action is low. On the other hand, a free institute performs better when the collective action's profitability is high. Furthermore, we show that in a structured population, when individuals have a choice between different institutes, a mutualistic relation between cooperators with different institute preferences emerges and helps the evolution of cooperation.
\end{abstract}
\section*{Introduction}
In many biological contexts, individuals in a group need to work collectively to solve collective action problems. Examples range from cooperation in resource acquisition in bacteria \cite{MacLean,Chen}, to cooperative defending or foraging \cite{MacLean,Smith,Pearson}, and cooperative breeding \cite{Holmes} in animal populations, and maintaining the commons such as fisheries \cite{Bodin}, or environmental preservation \cite{Barfuss,Bodin} in human societies. Successful solution of such collective action problems requires biological populations to find ways to suppress free-riding among the individuals. Past researches suggest formal and informal social institutions play an important role in solving such collective action problems. Informal social institutions can be at work, for instance, in the form of social norms \cite{Nowak,Fehr0,Nowak2,Takahashi,Brandt,Panchanathan,Leimar}, reputation and gossip \cite{Dunbar,Nowak,Wu,Hilbe0}, or even human language \cite{Salahshour0,Dunbar}. Such institutions can promote cooperation by channeling the benefit of cooperation towards cooperators. Human institutions, formal or informal, can promote cooperation through other mechanisms as well, such as reward \cite{Rand,Attila,Hilbe,Hu,Salahshour1}, and punishment \cite{Fehr,Perc,Boyd2,Szolnoki,Hauert1,Hilbe2,Salahshour2}. Such institutions can help to solve collective action problems by rewarding social behavior or punishing anti-social behavior.

Recently, it is also shown having a choice between different institutions to perform a collective action task can promote cooperation in the absence of any of such enforcement mechanisms \cite{Salahshour3}. Here, we bring attention to a relatively simple effect that can promote cooperation in costly institutes, when a choice between different institutes exist. We show that in a situation where individuals have a choice between different institutes, the existence of an entrance cost to enter an institute can keep free-riders away. In such a context, free-riders predominantly join the free institute. On the other hand, Cooperators, by working cooperatively in a costly institute, can obtain a higher profit despite paying an entrance cost that has no direct benefit. As our analysis shows, in both mixed and structured populations, a costly institute out-performs a free institute when the collective action's profitability is small. On the other hand, a free institute performs better when the profitability of the collective action is high. Furthermore, the existence of a costly alternative has a positive effect on the evolution of cooperation in a free institute.

Besides, we show that when a choice between different institutes exists, a synergistic mutualism between cooperators with different institute preferences emerges. This mutualistic relation among the cooperators, together with competition between defectors with different game preferences over scarce spatial resources, can help the evolution of cooperation in a spatial structure. Our findings can shed light on the solution of collective action problems and the function of collective action institutions in many circumstances where individuals can join or form different institutes to perform a collective action task.

\section*{The Model}
We consider a population of $N$ individuals. At each time step, groups of $g$ individuals are drawn at random from the population pool. Individuals in each group can choose between two public resources: A costly institute, which we call the resource $1$, and a free institute, which we call resource $2$. The costly institute has a participation fee. The participation fee is paid by all the individuals who choose this resource. Individuals gather payoff by playing a public goods game in their institute. In this game, individuals can either cooperate or defect. Cooperators pay a cost $c$ to invest the same amount in the public resource. Defectors pay no cost and do not invest. All the investments in a public resource $i$ is multiplied by an enhancement factor $r_i$ and is divided equally among the individuals in that resource. In addition to the public goods game, we assume individuals receive a base payoff, $\pi_0$, from other activities not related to the public goods game.

After playing the games, individuals reproduce with a probability proportional to their payoff. In the reproduction stage, the whole population is updated such that the population size remains constant. That is, for each individual in the next generation, an individual is chosen as a parent with a probability proportional to its payoff. The offspring inherit the game preference (which can be institute $1$ or $2$) and the game strategy (which can be cooperation $C$ or defection $D$) of its parent, subject to mutations. Mutation in the game preference and strategy occurs independently, each with probability $\nu$. In case a mutation occurs, the value of the corresponding variable is changed to its opposite value (for example, $1$ to $2$ for the game preference and $C$ to $D$ for the strategy).

In addition to a mixed population, we consider a structured population, in which the individuals reside on a network. We consider a first nearest neighbor square lattice with von Neumann connectivity and periodic boundaries for the population network. Each individual participates in five groups, each centered around itself or one of its neighbors, to perform a collective action task. Individuals in each group enter and play a public goods game in their preferred institute. The payoff of the individuals from the public goods game is defined as their average payoff from all the games that the individuals participates in. In addition to the payoff from the public goods games, individuals receive a base payoff $\pi_0$. After deriving their payoffs, the whole population is updated. For the reproduction, we consider an imitation or death-birth process in which each individual imitates the strategy of one of the individuals in its extended neighborhood, chosen with a probability proportional to its payoff. We assume mutations can occur as well. After imitation, the individual's strategy and game preference mutate independently and each with probability $\nu$.
\section*{Results}

\begin{figure}
	\centering
	\includegraphics[width=1\linewidth, trim = 41 262 35 57, clip,]{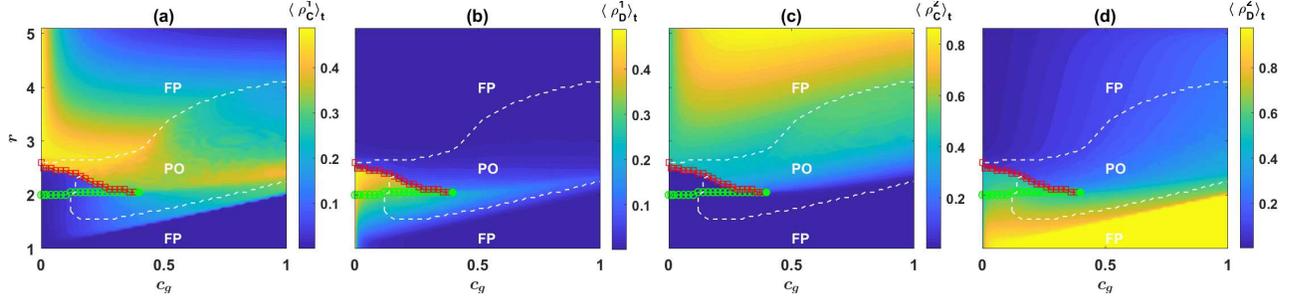}
	\caption{The density of different strategies in the $c_g-r$ plane. The densities of different strategies in the $c_g-r$ plane are color plotted. The phase diagram of the model is superimposed. For both small and large enhancement factors, $r$, the dynamics settle in a fixed point, denoted by FP. In between, the dynamics settle in a periodic orbit, denoted by PO. White lines show the boundary of the cyclic phase. For a small cost, the model is bistable for medium values of $r$. Green circles show the lower boundary of the bistable region, above which the cooperative orbit becomes stable. The red squares show the upper boundary of the bistable region above which the dynamics settle in the cooperative periodic orbit starting from all the initial conditions. The filled green circle shows the point where the transition between the two periodic orbits becomes a continuous transition. Parameter values: $g=5$, $nu=10^{-3}$, $\pi_0=2$. The replicator dynamic is solved for $8000$ time steps and time average are taken over the last $2000$ steps.}
	\label{figmixed1}
\end{figure}

\begin{figure}[!ht]
	\centering
	\includegraphics[width=1\linewidth, trim = 71 190 115 80, clip,]{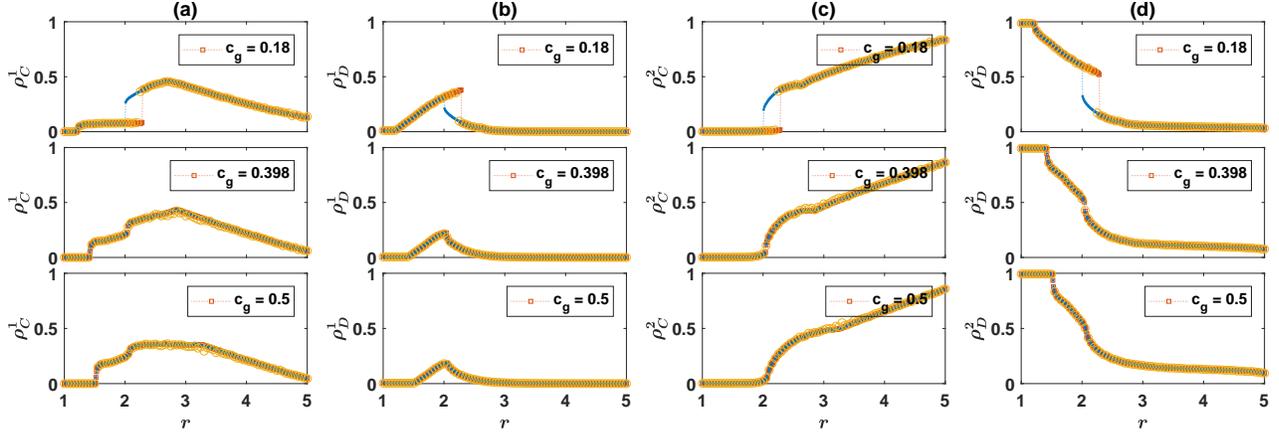}
	\caption{Density of different strategies as a function of $r$ for three different values of cost. The replicator dynamics is solved for two different initial conditions, a cooperation favoring initial condition in which all the individuals are cooperators and prefer the costly institute (blue dots), and a uniform initial condition in which strategy and game preference of the individuals are assigned at random (red squares). The result of simulations in a population of size $N=10000$ starting from a random initial condition is shown by orange circles. The system shows two different cooperative phases. For small enhancement factors, cooperation in the costly institute, but not in the free institute, evolves. For larger enhancement factors, cooperation in both costly and free institutes evolves. While for a small cost, the transition between the two cooperative phases is discontinuous and shows bistability (a), for high cost, there is a cross-over between the two phases by increasing the enhancement factor (c). Parameter values: $g=5$, $nu=10^{-3}$, $\pi_0=2$. The replicator dynamic is solved for $9000$ time steps, and the time averages are taken over the last $2000$ time steps. The simulation is performed for $6000$ time steps, and the averages are taken for the last $3000$ time steps.}
	\label{figmixed2}
\end{figure}

\subsection*{Mixed population}
As shown in the \nameref{Methods}, the model can be solved analytically in terms of the replicator-mutator dynamics in a mixed population. We begin, by setting $r_1=r_2=r$, and color plot the densities of different strategies in the $c_g-r$ plane, in Fig. \ref{figmixed1}. Here, the replicator dynamic is solved starting from a uniform initial condition in which all the strategies' initial density is equal. The phase diagram of the model is superimposed as well. The results of simulations in finite populations are in good agreement with the replicator dynamics results (see Fig. S1). We also plot the time average densities of different strategies as a function of $r$, for three different values of $c_g$, in Fig. \ref{figmixed2}. Blue dots and red squares represent the solutions of the replicator dynamics for two different initial conditions, and orange circles show the results of simulations starting from a uniform initial condition in which the individuals' strategies are assigned at random. Throughout this manuscript we fix $c=1$.

For small enhancement factors $r$, the dynamics settle in a fixed point where only defectors in the free institute survive. As $r$ increases beyond a threshold, the advantage of the costly institute becomes apparent: Cooperators in the costly institute, but not in the free institute, start to appear in the system. This shows paying a participation cost works as an incentive for the individuals to cooperate. However, as the enhancement factor increases, public goods' return can increase well above the participation cost. This gives an incentive for free-riding: In this region, free-riders' density in the costly institute increases slowly by increasing $r$.

\begin{figure}[!ht]
	\centering
	\includegraphics[width=1\linewidth, trim = 42 325 55 37, clip,]{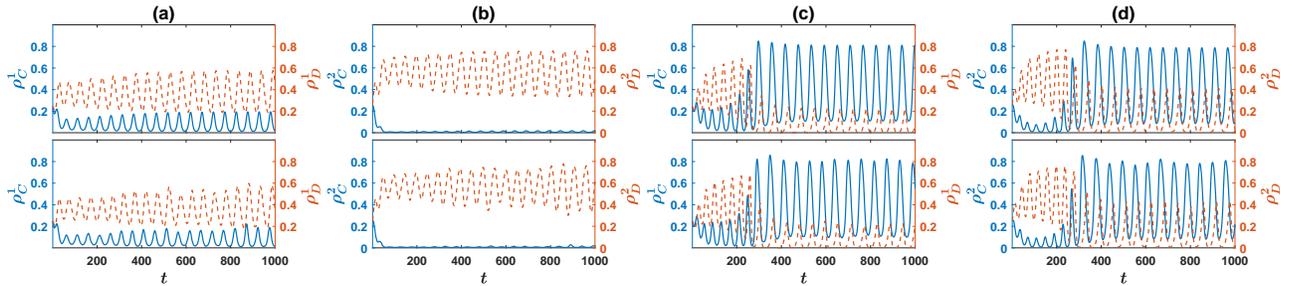}
	\caption{Examples of the time evolution of the system. Examples of the time evolution of the system for the defective periodic orbit, (a) and (b), and the cooperative periodic orbit (c) and (d). The top panels show the replicator dynamics results, and the bottom panels show the result of a simulation in a population of size $N=40000$ individuals. While in the defective periodic orbit only in the costly institute, cooperation evolves, in the cooperative periodic orbit, cooperation in both institutes evolves. Parameter values: $g=5$, $nu=10^{-3}$, $\pi_0=2$, and $c_g=0.18$. In (a) and (b) $r=2.2$, and in (c) and (d) $r=2.35$.}
	\label{figtime}
\end{figure}

For large enough costs, as $r$ increases, at a transition line indicated by the dashed white line in Fig. \ref{figmixed1}, temporal fluctuations sets in. These fluctuations are derived from the cyclic dominance of different strategies. The model shows two qualitatively different periodic orbits. In the first one, which occurs for smaller values of $r$, cooperation in the costly institute, but not in the free institute evolves. We call this periodic orbit the defective periodic orbit. For larger values of $r$, cooperation in both the costly and the free institutes evolves. We call this the cooperative periodic orbit, owing to the fact that cooperation in both institutes is viable in this regime. Finally, as $r$ increases beyond a final threshold (white dashed line for large values of $r$), the system settles in a cooperative fixed point, where most of the individuals prefer the free institute and cooperate.

For small costs, the system possesses a bistable region where both periodic orbits are stable. Green circles show the lower boundary of the bistable region, below which the cooperative orbit is unstable. Its upper boundary, above which the defective periodic orbit becomes unstable, is plotted by red squares, in Fig. \ref{figmixed1}.

The bistability of the dynamics for small costs can be seen in Fig. \ref{figmixed2} (top panels), as well. Here, the blue dots and red squares show the replicator dynamics' stationary state starting from two different initial conditions. While for both small and large $r$ the dynamic is mono-stable and the attractors of the dynamic for different initial conditions coincide, the situation is different for medium values of $r$: For $r$ between around $2$ and $3$, the system shows a bistable region where the two initial conditions result in different attractors. On the other hand, for large costs (bottom panels), no bistability is observed, and the stationary state of the dynamics is independent of the initial condition. In this region, by increasing $r$, the system shows a cross-over from the defective periodic orbit to the cooperative periodic orbit without passing any singularity. In between, the transition between the two periodic orbits becomes continuous at a single critical point (middle panel).

\begin{figure}
	\centering
	\includegraphics[width=1\linewidth, trim = 41 262 35 57, clip,]{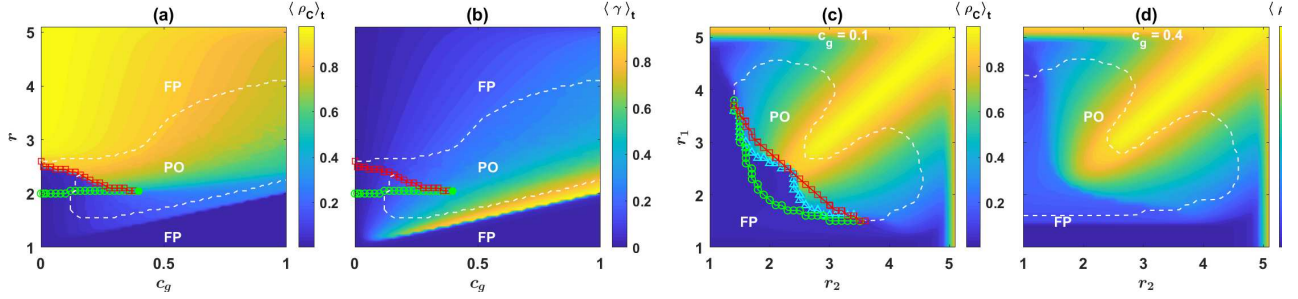}
	\caption{Evolution of cooperation. (a): Time average total density of cooperators, $\rho_C=\rho_C^1+\rho_C^2$ in the $r-c_g$ plane. (b): Time average difference between the probability that an individual in the costly institute is cooperator from the probability that an individual in the free institute is a cooperator, $\gamma=\rho_C^1/(\rho_C^1+\rho_D^1)-\rho_C^2/(\rho_C^2+\rho_D^2)$. Individuals are always more likely to be cooperator in the costly institute. (c) and (d): The time average total density of cooperators, $\rho_C=\rho_C^1+\rho_C^2$ in the $r_1-r_2$ plane for $c_g=0.1$ (c) and $c_g=0.4$ (d). Parameter values: $g=5$, $nu=10^{-3}$, and $\pi_0=2$. The replicator dynamics is solved for $8000$ time steps and time average are taken over the last $2000$ steps. }
	\label{figmixed3}
\end{figure}

\begin{figure}[!ht]
	\centering
	\includegraphics[width=1\linewidth, trim = 41 42 35 57, clip,]{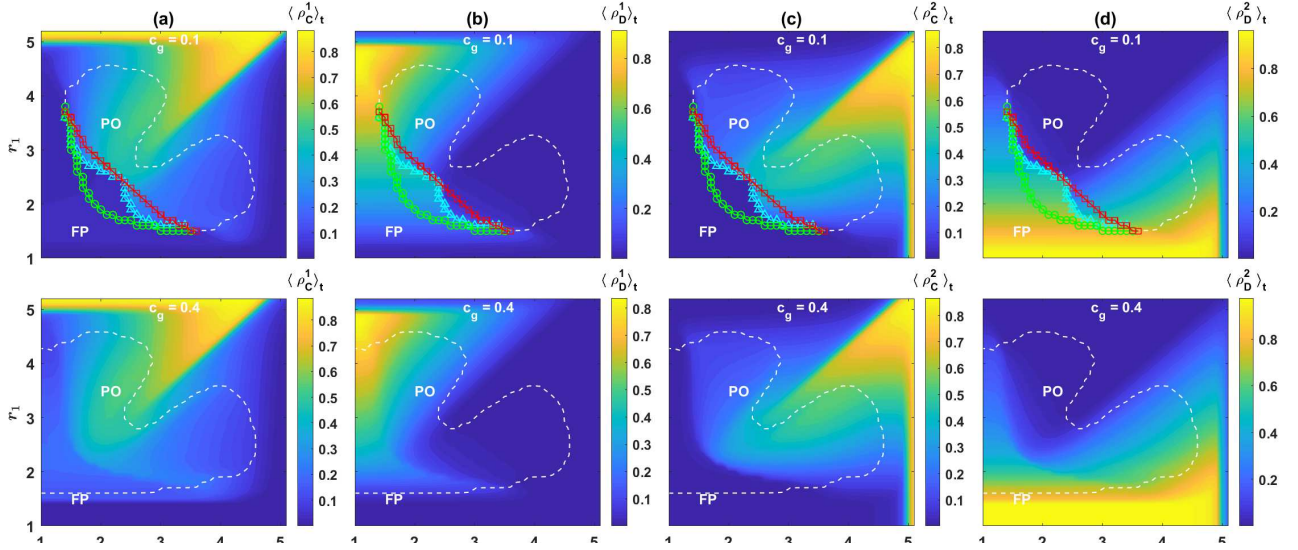}
	\caption{The density of different strategies in the $r_1-r_2$ plane. The densities of different strategies in the $r_1-r_2$ plane, for two different costs, $c_g=0.1$ (top) and $c_g=0.4$ (bottom), are color plotted. The phase diagram of the model is superimposed. For a small cost ($c_g=0.1$, top), for both small and large enhancement factor values, $r$, the dynamics settle in a fixed point, denoted by FP. In between, the dynamics settle in a periodic orbit, denoted by PO. White lines show the boundary of the cyclic phases. For a small cost, the model is bistable for medium values of $r$. Green circles show the lower boundary of the bistable region, above which the cooperative periodic orbit becomes stable. The red squares show the upper boundary of the bistable region above which the dynamics settle in the cooperative periodic orbit starting from all the initial conditions. Blue triangles show the phase boundary, resulting from a uniform initial condition. For a large cost ($c_g=0.4$, bottom), the bistability is lost, and the periodic phase's domain increases. Parameter values: $g=5$, $nu=10^{-3}$, and $\pi_0=2$. The replicator dynamic is solved for $8000$ time steps and time average are taken over the last $2000$ steps.}
	\label{figmixedr1r2}
\end{figure}

An example of the defective periodic orbit, in smaller values of $r$, is presented in Figs. \ref{figtime}(a) and \ref{figtime}(b). Here, $r=2.2$ and $c_g=0.18$. The top panels show the replicator dynamics results, and the bottom panels show the result of a simulation in a population of $N=40000$ individuals. $D^1$ experiences an advantage over $C^1$, and proliferates when the density of $C^1$ increases. Being exploited by $D^1$, the density of $C^1$ declines when $D^1$ increases in the population. This decreases the return in the costly institute below its entrance cost. At this point, due to not paying an entrance cost, $D^2$ performs better than $D^1$ and increases in density.

The dynamic of the model is different in the cooperative periodic orbit. This can be seen in Fig. \ref{figtime}(c) and \ref{figtime}(d), where an example of the cooperative periodic orbit is presented. Here, $r=2.35$ and $c_g=0.18$. The top panels show the replicator dynamics results, and the bottom panels show the result of a simulation in a population of $N=40000$ individuals. In this case, cooperation in both institutes evolves. As cooperators' density in an institute $i$ increases, defectors in this institute start to increase in number. This decreases the profitability of the institute $i$, and thus, cooperators in the competing institute increase in number. This, in turn, motivates free-riding in the competing institute. In this way, the cyclic dominance of the four possible strategies, derives periodic fluctuations in the time evolution of the system.

An interesting question is how the population's cooperation depends on the participation cost and the enhancement factors. To see this, in Fig. \ref{figmixed3}(a) we plot the density of cooperators in the population, $\rho_C=\rho_C^1+\rho_C^2$. For large $r$, increasing the entrance cost of the costly institute has a detrimental effect on cooperation in both institutes. This is because, although the density of defectors in the costly institute remains close to zero, fewer individuals are willing to choose a costly institute with a high cost. This decreases the density of costly cooperators in the population. Due to its inability to attract individuals, the costly institute plays a less viable role as an alternative to a free institute. As having such a choice between different institutes promotes cooperation \cite{Salahshour3}, the limitation imposed on the individuals' choice by a high participation cost can deteriorate cooperation in the free institute, as well. 

On the other hand, having a participation cost can be beneficial for small enhancement factors. This is because an entrance cost deters defectors from entering the costly institute. To more clearly see this is the case, we consider the difference between the probabilities that an individual in the costly institute is a cooperator and the probability that an individual in the free institute is a cooperator, $\gamma=\rho_C^1/(\rho_C^1+\rho_D^1)-\rho_C^2/(\rho_C^2+\rho_D^2)$. This is contour plotted in Fig. \ref{figmixed3}(b), where it can be seen it is always positive. This fact increases a costly institute's profitability and allows individuals to reach a higher payoff by entering the costly institute.

So far, we have assumed that the costly institute and the free institute have the same quality. In general, the quality of the two resources may be different. To study this case, in Figs. \ref{figmixed3}(c) and \ref{figmixed3}(b) we plot the density of cooperators in the population, $\rho_C=\rho_C^1+\rho_C^2$, in the $r_1-r_2$ plane, for two different costs. The phase diagram of the model is superimposed as well. The density of different strategies in these cases are plotted in Fig. \ref{figmixedr1r2} (See Figs. S2 and S3 in the Supplemental Information for comparison with simulations). In the case of a small cost, $c_g=0.1$, the model settles in a fixed point with a low level of cooperation for small enhancement factors. As the enhancement factors increase, the model becomes bistable: In addition to the fixed point, a periodic orbit in which cooperators survive and cyclically dominate the population emerges. Green circles plot the lower boundary of the bistable region, and its upper boundary, above which the fixed point becomes unstable, is plotted by red squares. The phase boundary can be defined as the boundary where a transition between the two phases occurs starting from a uniform initial condition in which the initial density of all the strategies are equal \cite{Binder}. This is marked by blue triangles. Further increasing the enhancement factors, above a final transition line plotted by the dashed white line, the dynamics settle in a fixed point where cooperators dominate the population. 

The situation is different for large cost ($c_g=0.4$). In this case, the model is mono-stable in the whole $r_1-r_2$ plane. In addition, even for small $r_2$ ($r_2=1$), cooperation in the costly institute can evolve as long as $r_1$ is larger than a rather small threshold (around $1.5$ in the figure). In fact, cooperation in the costly institute evolves even when $r_2<1$: As individuals in the free institute defect for too small $r_2$ and the free institute yields a zeros outcome, decreasing the value of $r_2$ below $1$ does not affect the evolution of cooperation in the costly institute. This shows that the existence of a viable alternative is not necessary for the evolution of cooperation in a costly institute. Rather, having an entrance cost acts as a deterrence mechanism that can control free-riding in the institute, as long as participation in the public goods is voluntary. This contrasts the situation in the free institute, where, for too small $r_1$, cooperation does not evolve even for large values of $r_2$. This, in turn, shows, maintenance of cooperation in a free institute is contingent on the presence of a viable alternative, which guarantees individuals to have a choice between different viable institutes.

\begin{figure}
	\centering	\includegraphics[width=1\linewidth, trim = 41 262 35 57, clip,]{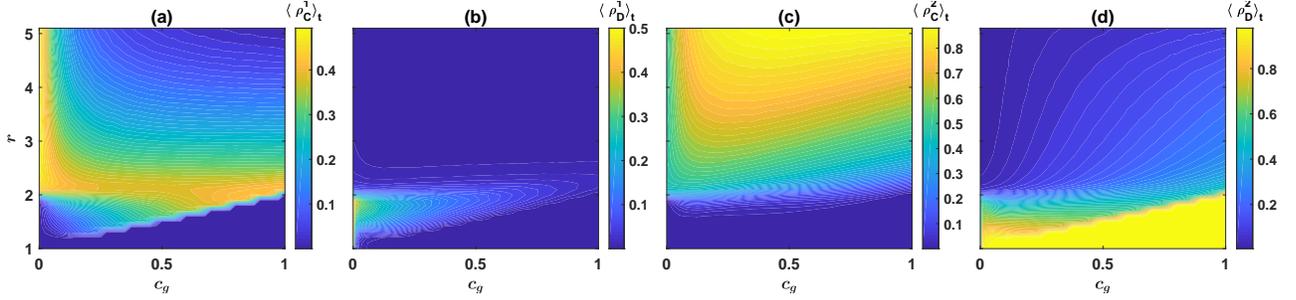}
	\caption{The density of different strategies in the $c_g-r$ plane in a structured population. The densities of different strategies in the $c_g-r$ plane are color plotted. Parameter values: $g=5$, $nu=10^{-3}$, and $\pi_0=2$. The population resides on a $200\times 200$ first nearest neighbor square lattice with von Neumann connectivity and periodic boundaries. The simulation is performed for $5000$ time steps starting from an initial condition in which all the individuals are defectors and prefer one of the two institutes at random. The time average is taken over the last $2000$ steps.}
	\label{fignetcg}
\end{figure}

\begin{figure}
	\centering
	\includegraphics[width=1\linewidth, trim = 41 42 35 42, clip,]{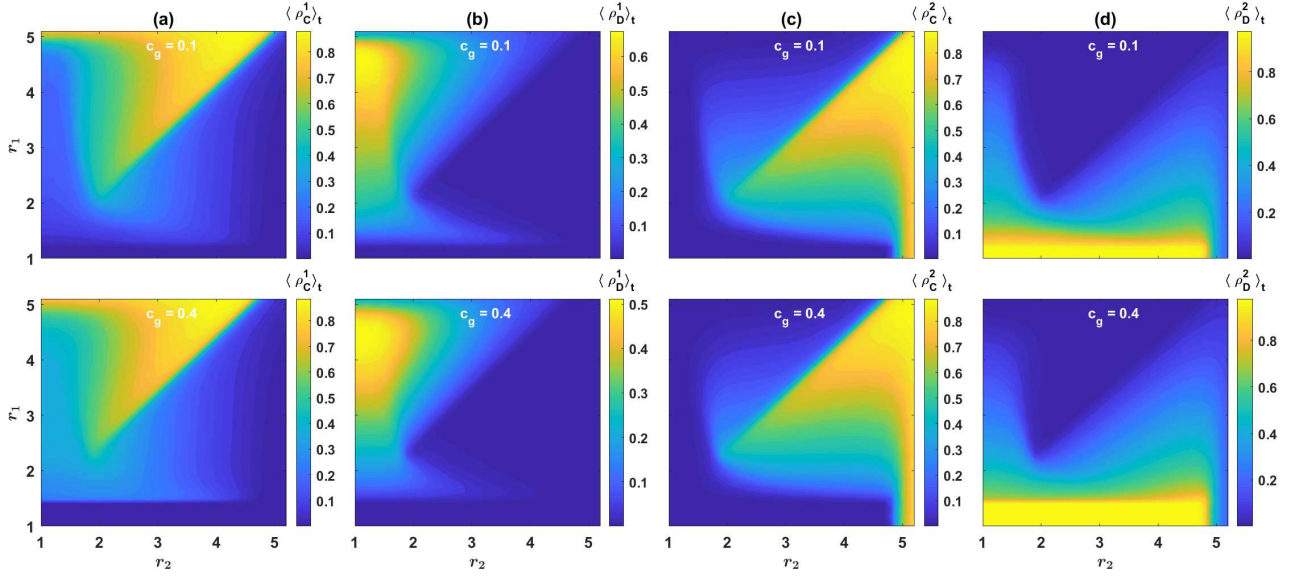}
	\caption{The density of different strategies in the $r_1-r_2$ plane in a structured population. The densities of different strategies in a structured population for two different costs, $c_g=0.1$ (top) and $c_g=0.4$ (bottom) are color plotted in the $r_1-r_2$ plane. The population resides on a $200\times 200$ first nearest neighbor square lattice with von Neumann connectivity and periodic boundaries. Parameter values: $g=5$, $nu=10^{-3}$, and $\pi_0=2$. The simulation is performed for $5000$ time steps starting from an initial condition in which all the individuals are defectors and prefer one of the two institutes at random. The time average is taken over the last $2000$ steps.}
	\label{fignetr}
\end{figure}

\subsection*{Structured population}
In contrast to the mixed population, the model shows no bistability in a structured population, and the fate of the dynamics is independent of the initial condition. To see why this is the case, we note that in a mixed population, a situation where all the individuals are defectors, and randomly prefer one of the two institutes, is the worst case for the evolution of cooperation, as in this case, mutant cooperators are in a disadvantage in both institutes. However, in a structured population, starting from such an initial condition, due to spatial fluctuations blocks of defectors, the majority of whom prefer the same institute form. A mutant cooperator who prefers the minority institute in its neighborhood obtains a high payoff and proliferates. This removes the bistability of the dynamics in a structured population. 

To study the model's behavior in a structured population, we perform simulations starting from an initial condition in which all the individuals are defectors and prefer one of the two institutes at random. The model shows similar behavior in a structured population to that in a mixed population. This can be seen in Fig. \ref{fignetcg} where the densities of different strategies are color plotted in the $c_g-r$ plane. As was the case in a mixed population, cooperation does not evolve for too small values of $r$. As $r$ increases beyond a threshold, cooperation does evolve in the costly institute, but not in the free institute. In this region, for a fixed enhancement factor, an optimal cost exists, which optimizes the cooperation level in the population. On the other hand, cooperation in both the costly and the free institutes evolves for large enhancement factors. In this region, increasing the cost can slightly increases defection in the free institute and have a detrimental effect on the evolution of cooperation, but not as much as it does in a mixed population.

To look at the model's behavior in a structured population when the two resources have different qualities, in Fig. \ref{fignetr} we plot the densities of different strategies in the $r_1-r_2$ plane for two different participation costs. In the top panels $c_g=0.1$, and in the bottom panels $c_g=0.4$. As mentioned before, contrary to a mixed population, the model does not show bistability in a structured population. Consequently, the behavior of the model is qualitatively similar for small and large costs. For too small $r_1$, for all the values of $r_2$, only non-costly defectors, $D^2$ survive. On the other hand, for $r_1$ larger than a small value (approximately $r_1=1.5$) costly cooperators survive even for $r_2$ equal or smaller than $1$. This shows, similarly to the case of a mixed population, having a choice between different resources is not necessary to promote cooperation in a costly institute. Rather, cooperation evolves in a costly institute even for relatively small enhancement factors as long as participation is optional. Increasing $r_2$ however increases cooperation level in the costly institute (as long as $r_2<r_1$). This results from the beneficial effect of freedom of choice between different public resources for the evolution of cooperation. Comparing the results for a small and a high participation cost shows that higher participation cost improves cooperation level in the costly institute for small $r_2$, i.e., when a viable alternative does not exist.

\begin{figure}
	\centering
	\includegraphics[width=1\linewidth, trim = 44 1 50 17, clip,]{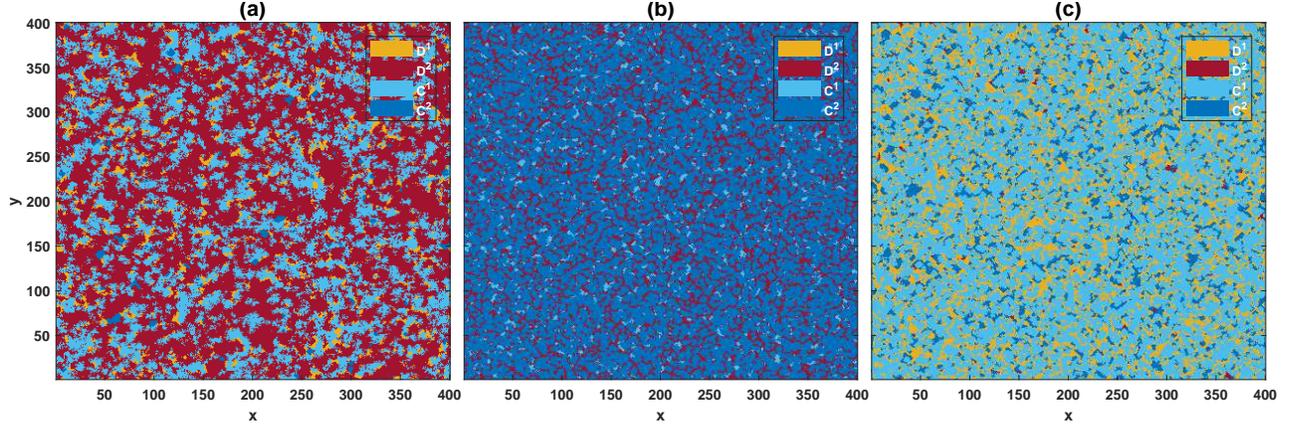}
	\caption{Snapshots of the population in the stationary state for different parameter values. Different strategies are color codded (legend). In (a), $r_1=1.7$, $1_2=1.7$, in (b), $r_1=3.5$ and $r_2=3.5$, and in (c), $r_1=3$, and $r_2=1.8$. In all the cases $c_g=0.6$. Here, individuals reproduce with a probability proportional to the exponential of their payoff with a selection parameter equal to $\beta=5$. The population resides on a $400\times400$ square lattice with von Neumann connectivity and periodic boundaries. Parameter values: $g=5$ and $\nu=10^{-3}$}
	\label{figsnapshot}
\end{figure}

On a spatial structure, the model's dynamic is governed by the cyclic dominance of different strategies, which can give rise to traveling waves. To take a deeper look into the dynamics of the system, in Fig. \ref{figsnapshot} we present snapshots of the population's stationary state in different phases. In this figure, we consider a model in which individuals reproduce with a probability proportional to the exponential of their payoff, $\pi$, times a selection parameter, $\beta$, $\exp(\beta\pi)$ (see the Supplementary Information S.1), with $\beta=5$. The situation in the model where individuals reproduce with a probability proportional to their payoff is similar. In Fig. \ref{figsnapshot}(a), we have set $r_1=r_2=1.7$. This phase corresponds to the defective periodic orbit in the mixed population case. Here the majority of the population are non-costly defectors. Costly cooperators experience advantage over the former and can proliferate in the sea of non-costly defectors. However, costly cooperators are at a disadvantage in comparison to both costly defectors and non-costly cooperators. The former can only survive in small bands around costly cooperators, as they rapidly get replaced by non-costly defectors once they eliminate costly cooperators. This phenomenon shows that competition between defectors with differing institute preferences can positively impact the evolution of cooperation. Non-costly cooperators, in turn, can survive by forming compact domains where they reap the benefit of cooperation among themselves. However, as in this region, the effect of network reciprocity is too small to promote cooperation, non-costly cooperators get eliminated by non-costly defectors once costly cooperators are out of the picture. Consequently, the dynamic of the system is governed by traveling waves of costly cooperators followed by small trails of costly defectors and non-costly cooperators in a sea of non-costly defectors (see the Supplementary Video, SV.1, for an illustration of the dynamics).

Fig. \ref{figsnapshot}(b), shows a snapshot of the population for $r_1=r_2=3.5$. In this region, non-costly cooperators dominate the population. However, non-costly defectors can survive in small bands in the sea of non-costly cooperators. While at a disadvantage in the sea of non-costly cooperators, costly cooperators beat non-costly defectors. Consequently, small blocks of costly cooperators are formed within the bands of non-costly defectors. These blocks of costly cooperators move along the bands of non-costly defectors and purge the population from non-costly defectors. In this way, although costly cooperators exist only in small density, they play a constructive role in helping non-costly cooperators to dominate the population.

Finally, a similar phenomenon can also occur in the opposite regime of $r_1>2_2$, where costly cooperators dominate the population. An example of this situation is plotted in Fig. \ref{figsnapshot}(c). Here, $r_1=3$, $r_2=1.8$ and $c_g=0.6$. In this case, costly cooperators dominate the population. Costly defectors can survive in the sea of costly cooperators in small bands. Non-costly cooperators, while at a disadvantage in the sea of costly cooperators, can grow within the small bands of costly defectors. A similar phenomenon holds for non-costly defectors who, due to not paying the participation cost, receive a higher payoff than costly defectors once costly cooperators are eliminated in their neighborhood. Both these latter types can form and grow in domains of costly defectors. This, in turn, helps the evolution of cooperation in the costly institute.

Analysis of the spatial patterns at work in the evolution of the system reveals competition or synergistic relation between individuals with different institute preference plays an essential role in the evolution of cooperation in the system. Defectors with different institute preferences always appear as competitors who compete over space. By eliminating each other, they play a surprisingly constructive role in the evolution of cooperation (this can be seen, for instance, in Fig. \ref{figsnapshot}(c)). Cooperators, on the other hand, while having a direct competition over scarce sites, can also act synergistically and help the evolution of cooperation in their opposite institute. In a neighborhood of defectors who prefer the same institute, a mutant cooperator who prefers the opposite institute performs better than the neighboring defectors and can grow. In this way, by purging defectors with an opposite game preference, cooperators help fellow cooperators with an opposite game preference. Consequently, cooperators with different game preferences, despite having direct competition for scarce spatial resources, can engage in a mutualistic relation and help each other to overcome defectors.

\section*{Discussion}

The question that how participation cost affects the evolution of cooperation had been considered in the context of binary interactions (prisoner's dilemma), for instance, in the context of indirect reciprocity \cite{Pena}, structured populations \cite{Masuda,Tanimoto,Li}, and in repeated group interactions \cite{Wang}. Some previous work show when interactions are obligatory, participation cost can have a detrimental effect on the evolution of cooperation \cite{Pena,Masuda}. As we have shown, the advantage of a costly institute becomes apparent when it is accompanied by an alternative free institute. In such contexts, while defectors predominantly join a free institute, cooperators can reach a higher payoff by working cooperatively in a costly institute, despite paying a participation cost with no direct benefit. Furthermore, the existence of a costly institute has a positive impact on the evolution of cooperation in the free institute. In a mixed population, this is brought about by the fact that non-costly defectors, those who join a free institute and free ride, can be eliminated by costly cooperators, those who join a costly institute and cooperate. While this mechanism can be at work to promote cooperation in a mixed population, it is further strengthened in the presence of population structure. In a structured population, costly cooperators, while at a disadvantage in the presence of non-costly cooperators or costly defectors, can beat non-costly defectors. Consequently, they can grow in domains of non-costly defectors, and by purging the population from non-costly defectors, they help the evolution of non-costly cooperators.

Analysis of the model in a structured population reveals interesting insights about how having a choice between different resources can help the evolution of cooperation. In a structured population, defectors with different game preferences are in direct competition over scarce spatial resources. By undermining each other, they help the evolution of cooperation in their opposite institute. Cooperators with opposite game preferences are in direct competition as well. However, a mutualistic relation emerges between these two as well. This comes about because cooperators can beat defectors with the opposite game preference and grow in their domain. By eliminating free-riders in the opposite institute, they help their rival cooperators with a different game preference to flourish. Consequently, a mutualistic relation between cooperators with opposite game preference emerges, which helps the evolution of cooperation in the population. We note that such a dynamic is at work to help the evolution of cooperation in a structured population in a more general context where individuals have a choice between different institutes, irrespective of the existence of a participation cost.

Empirical examples of a situation where individuals have a choice between different public resources to perform a collective action task appear to abound in human societies. Some examples include competition between political parties, firms, and associations, team incentives within firms \cite{Sheremeta}. A choice between geographic locations such as cities \cite{Ziblatt,Batabyal}, or having a choice between different collective action tasks such as communal hunting or communal agriculture \cite{Campana} provide other examples. As our analysis suggests, the existence of a costly alternative can positively impact the evolution of cooperation both in the costly resource and its competing resources. In the case of large scale human institutes, such a participation cost can be costly signaling of a common goal, for example, in the form of a donation to charity \cite{Jones}, or transaction cost associated with the search, bargaining, or monitoring to maintaining the commons \cite{Taylor}. Formation and maintenance of costly relations \cite{Hanaki,Bednarik,Masuda}, when some heterogeneity in formation and maintenance costs exists, can be another example at work in a situation where individuals can form or join different groups with different participation costs, to perform a collective action task.

Examples of a situation where individuals have a choice between different resources appear to exist in other biological populations as well. An example is the fission-fusion dynamics observed in different animal groups \cite{Smith,Holmes,Aureli}, or even hunter-gatherer human societies \cite{Marlowe}. In many such cases, sub-groups are formed within a group to perform a collective action task, such a cooperative hunting \cite{Smith,Pearson}, or cooperative defending \cite{Smith}. Another example is resource competition in microorganisms \cite{Chen,MacLean}. Our analysis predicts cooperation is easier to evolve when joining a group or embarking on a collective action task is more costly than others. Future empirical research can shed more light on the extent to which such a mechanism is at work in different biological populations.

\section*{Methods}
\label{Methods}
\subsection*{The replicator dynamics}
The model can be described in terms of the replicator-mutators equation \cite{Nowak3}. This equation reads as follows:
\begin{align}
\rho_x^i(t+1)=\sum_{y,j}\nu_{y,j}^{x,i}\rho_y^j(t)\frac{\pi_y^j(t)}{\sum_{z,l}\rho_z^l(t)\pi_z^l(t)}.
\label{eqrep}
\end{align}
Here, $x$, $y$, and $z$ refer to strategies and can be either $C$ or $D$, and $i$, $j$, and $l$ refer to the public resources which can be $1$ or $2$. $\nu^{x,i}_{y,j}$ is the mutation rate from a strategy profile that prefers public resource $j$ and plays strategy $y$ to a strategy combination that prefers public resource $i$ and plays strategy $x$. These can be written in terms of mutation rates as follows:
\begin{align}
\begin{cases}
\nu^{x,i}_{y,j}=1-2\nu+\nu^2, \quad\quad &\textit{if\quad ($i=j$\quad and \quad $x=y$)}\\
\nu^{x,i}_{y,j}=\nu-\nu^2, \quad\quad& \textit{if\quad ($i=j$\quad and \quad $x\neq y$)}\quad \textit{or}\quad\textit{($i\neq j$\quad and \quad $x=y$)}\\
\nu^{x,i}_{y,j}=\nu^2\quad\quad&\textit{if\quad($i\neq j$\quad and \quad $x\neq y$)}
\end{cases}
\end{align}
In eq. (\ref{eqrep}), $\pi_y^j$ is the expected payoff of an individual who prefers public resource $j$ and plays strategy $y$. These terms can be written by averaging a focal individual's payoff with game preference $j$ (for $j=1$ and $2$) in a group composed of $n_C^j$ cooperators and $n_D^j$ defectors who prefer public resource $j$, over all possible group configurations. In this way, we have the following equations for the payoffs:
\begin{align}
\pi_C^1=&\sum_{n_D^1=0}^{g-1-n_C^1}\sum_{n_C^1=0}^{g-1}cr_1\frac{1+n_C^1}{1+n_C^1+n_D^1}\nonumber\\&(1-\rho_C^1-\rho_D^1)^{g-1-n_C^1-n_D^1}{\rho_D^1}^{n_D^1}{\rho_C^1}^{n_C^1}\binom{g-1}{n_C^1,n_D^1,g-1-n_C^1-n_D^1}-c-c_g+\pi_0,\nonumber\\
\pi_D^1=&\sum_{n_D^1=0}^{g-1-n_C^1}\sum_{n_C^1=0}^{g-1}cr_1\frac{n_C^1}{1+n_C^1+n_D^1}\nonumber\\&(1-\rho_C^1-\rho_D^1)^{g-1-n_C^1-n_D^1}{\rho_D^1}^{n_D^1}{\rho_C^1}^{n_C^1}\binom{g-1}{n_C^1,n_D^1,g-1-n_C^1-n_D^1}-c_g+\pi_0.\nonumber
\end{align}
\begin{align}
\pi_C^2=&\sum_{n_D^2=0}^{g-1-n_C^2}\sum_{n_C^2=0}^{g-1}cr_2\frac{1+n_C^2}{1+n_C^2+n_D^2}\nonumber\\&(1-\rho_C^2-\rho_D^2)^{g-1-n_C^2-n_D^2}{\rho_D^2}^{n_D^2}{\rho_C^2}^{n_C^2}\binom{g-1}{n_C^2,n_D^2,g-1-n_C^2-n_D^2}-c+\pi_0,\nonumber\\
\pi_D^2=&\sum_{n_D^2=0}^{g-1-n_C^2}\sum_{n_C^2=0}^{g-1}cr_2\frac{n_C^2}{1+n_C^2+n_D^2}\nonumber\\&(1-\rho_C^2-\rho_D^2)^{g-1-n_C^2-n_D^2}{\rho_D^2}^{n_D^2}{\rho_C^2}^{n_C^2}\binom{g-1}{n_C^2,n_D^2,g-1-n_C^2-n_D^2}+\pi_0.
\end{align}
In this equation, $c r_1\frac{1+n_C^1}{1+n_C^1+n_D^1}-c$ in the first equation is the payoff of a cooperator who prefers the public resource $1$, and $cr_1\frac{n_C^1}{1+n_C^1+n_D^1}$ in the second equation is the payoff of a defector who prefers public resource $1$. $(1-\rho_C^1-\rho_D^1)^{g-1-n_C^1-n_D^1}{\rho_D^1}^{n_D^1}{\rho_C^j}^{n_C^1}\binom{g-1}{n_c^1,n_D^1,g-1-n_C^1-n_D^1}$, is the probability that such a group composition occurs. Here, $\binom{g-1}{n_c^1,n_D^1,g-1-n_C^1-n_D^1}=\frac{(g-1)!}{n_C^1!n_D^1!(g-1-n_C^1-n_D^1)!}$ is the multinomial coefficient and is the number of ways that $n_C^1$ cooperators and $n_D^1$ defectors who prefer game $1$ can be chosen among $g-1$ group-mates of a focal individual. Summation over all the possible configurations gives the expected payoff of a cooperator, or defector who prefers resource 1. As resource one is a costly institute, the individual needs to pay a participation cost, $c_g$. Finally, a base payoff of $b$ is added to all the payoffs.

It is easy to derive the expected payoff of those who prefer public resource $2$, using a similar argument.

\subsection*{Solution of the replicator dynamics and simulations}
Figs. \ref{figmixed1}, \ref{figmixed3}, and Fig. \ref{figmixedr1r2} result from numerical solution of the replicator dynamics. The replicator dynamics is solved for $T=8000$ time steps, and an average over the last $2000$ time steps is taken. The initial condition is a uniform initial condition in which all the strategies' initial densities are equal ($\rho_C^1=\rho_D^1=\rho_D^1=\rho_D^2=0.25$). The simulations presented in Figs. \ref{figmixed2} and \ref{figtime} are performed starting from a random assignment of the strategies. In Fig. \ref{figmixed2} the simulations are performed for $6000$ time steps, and an average over the last $3000$ time steps is performed. The results of the replicator dynamics in Fig. \ref{figmixed2} result from numerical solutions of the replicator dynamics starting from two different initial conditions. The blue dots show the results starting from a cooperation favoring initial condition in which at the beginning of the simulation, all the individuals are cooperators and prefer the costly institute. The red squares show the results starting from a uniform initial condition. The simulations in a structured population presented in Figs \ref{fignetcg} and \ref{fignetr} are performed for $5000$ time steps starting from an initial condition in which all the individuals are defectors and prefer one of the two institutes chosen randomly. The time averages are taken over the last $2000$ time steps. As the model shows no bistability in a structured population, different initial conditions result in the same picture.

To derive the phase diagrams, we solve the replicator dynamics starting from different initial conditions. In the mono-stable region, the stationary state of the dynamics is the same for different initial conditions. This allows us to derive the boundary of the cyclic phase (the white line) using any of the initial conditions. To determine the boundary of bistability, we note the most facilitative initial condition for the evolution of cooperation is the one in which all the individuals are cooperators and prefer the costly institute. The lower boundary of the bistable region above which the cooperators can survive is determined starting from this initial condition. The worst initial condition for the evolution of cooperation is the one in which all the individuals are defectors and prefer the two institutes in similar frequencies. The replicator dynamics solutions starting from this initial condition give the upper boundary of the bistable region above which the defective state becomes unstable.

\section*{Acknowledgments}
The author acknowledges funding from Alexander von Humboldt Foundation in the framework of the Sofja Kovalevskaja Award endowed by the German Federal Ministry of Education and Research. 


\end{document}